\documentclass[aps,prb,twocolumn,superscriptaddress,showpacs]{revtex4-1}

\usepackage{amsmath}
\usepackage{graphicx}
\usepackage{calc}
\usepackage{bm}

\usepackage[T2A]{fontenc}
\usepackage[cp1251]{inputenc}
\usepackage[english]{babel}

\begin{document}

\title{Evidence of the ferroelectric polarization in charge transport through WTe$_2$ Weyl semimetal surface}

\author{N.N. Orlova}
\author{N.S. Ryshkov}
\author{A.V. Timonina}
\author{N.N. Kolesnikov}
\author{E.V. Deviatov}

\affiliation{Institute of Solid State Physics of the Russian Academy of Sciences, Chernogolovka, Moscow District, 2 Academician Ossipyan str., 142432 Russia}

\date{\today}

\begin{abstract}

We investigate electron transport along the surface of WTe$_2$ three-dimensional single crystals, which are characterized by coexistence of Weyl semimetal conductivity and ferroelectricity at room temperature. We find that non-linear behavior of $dV/dI(I)$ WTe$_2$ differential resistance is accompanied by slow relaxation process, which appears as the $dV/dI(I)$ dependence on the sign of the current change. This observation is confirmed by direct investigation of time-dependent relaxation curves. While strongly non-linear  differential resistance  should be expected for the zero-gap WTe$_2$, the slow relaxation  in  transport is very unusual for well-conducting semimetals  at room temperature. We establish, that non-monotonous dependence of the amplitude of the effect on driving current $\Delta dV/dI(I)$  well corresponds to  the known Sawyer-Tower's  ferroelectric hysteresis loop. This conclusion is also confirmed by gate voltage dependencies, so our results can be understood as a direct demonstration of WTe$_2$ ferroelectric polarization in charge transport experiment.

\end{abstract}


\maketitle

\section{Introduction}

The layered transition metal dichalcogenides (TMDCs) are attracting significant research interest due to their extraordinary electronic and optical properties. For instance, band gap varies  from indirect to direct one in single layers of MoS$_2$, MoSe$_2$, WS$_2$ and WSe$_2$, which allows different possible applications like  transistors, photodetectors and electroluminescent devices~\cite{MoS2,WS2,rewievMoS2_WS2}. The recent discovery of the memristor effect (or nonvolatile resistance switching) in monolayers TMDC structures has added a new field of fundamental and applied activities~\cite{NVRS}. For MoS$_2$ monolayers,  the  effect is based on the  gold adatom absorption at native vacancy defects~\cite{NVRS1}. 

WTe$_2$ is a special member among the TMDCs family. It has been extensively studied~\cite{II-WSM1,II-WSM2,II-WSM3,II-WSM4} as realization of a  Weyl semimetal  with broken inversion symmetry for three-dimensional samples~\cite{WSM1,WSM2,WSM3}, while monolayer WTe$_2$ is known to be quantum spin Hall insulator~\cite{QSH1,QSH}. In ambient conditions WTe$_2$ is characterized by phase distorted 1T' (Td) crystal structure in a wide temperature range~\cite{WTe2_str1,WTe2_str2,WTe2_str3},  in contrast to typical TMDCs, where several (2H or 1T) phases can be observed~\cite{structure}. At low temperatures, WTe$_2$ is characterized by extremely large unsaturated magnetoresistance~\cite{MR} and even superconductivity~\cite{SC} at high pressures.

Recently, three-dimensional WTe$_2$ single crystals were found to demonstrate  coexistence of metallic conductivity and ferroelectricity at room temperature~\cite{WTe2_fer}. The latter usually belongs to the insulators, but it occurs in WTe$_2$ due to the strong anisotropy of the non-centrosymmetric crystal structure.  The spontaneous polarization of ferroelectric domains is found to be bistable, it can be affected by high external electric field~\cite{WTe2_fer}. Scattering of the charge carriers on the domain walls is known to provide noticeable contribution to the sample resistance~\cite{domain wall effect on transp}. Thus, coexistence of metallic and ferroelectric properties should produce new physical effects~\cite{polar} for electron transport in TMDCs, and, therefore, it should  be important for nanoelectronic applications.

Here, we investigate electron transport along the surface of WTe$_2$ three-dimensional single crystals, which are characterized by coexistence of Weyl semimetal conductivity and ferroelectricity at room temperature. We find that non-linear behavior of $dV/dI(I)$ WTe$_2$ differential resistance is accompanied by slow relaxation process, which appears as the $dV/dI(I)$ dependence on the sign of the current change. This observation is confirmed by direct investigation of time-dependent relaxation curves. While strongly non-linear  differential resistance  should be expected for the zero-gap WTe$_2$, the slow relaxation  in  transport is very unusual for well-conducting semimetals  at room temperature. We establish, that non-monotonous dependence of the amplitude of the effect on driving current $\Delta dV/dI(I)$  well corresponds to  the known Sawyer-Tower's  ferroelectric hysteresis loop. This conclusion is also confirmed by gate voltage dependencies, so our results can be understood as a direct demonstration of WTe$_2$ ferroelectric polarization in charge transport experiment.

\section{Samples and techniques}

\begin{figure}[t]
\center{\includegraphics[width=\columnwidth]{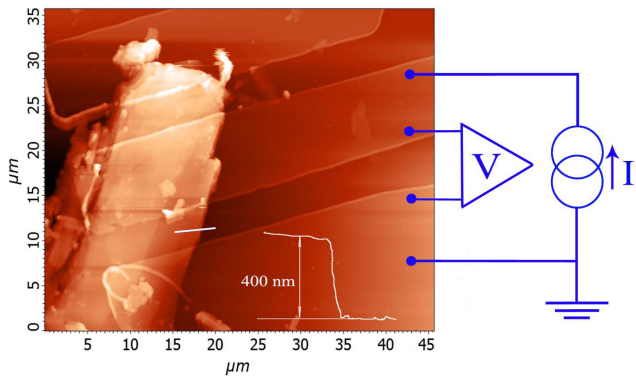}}
\caption{ AFM image of the 400~nm thick sample with Au leads, evaporated over the WTe$_2$ flake. The leads are separated by 5~$\mu$m intervals. Inset demonstrates the AFM scan of the flake profile between the contact leads, along the white line in the image.  Schematic diagram of the  measurement circuit is also shown for the two-point connection scheme. For correct measurement of the low-resistance samples, we define current $I$ between the leads  and measure the resulting voltage drop $V$.}
\label{sample}
\end{figure}

WTe$_2$ compound was synthesized from elements by reaction of metal with tellurium vapor in the sealed silica ampule. The WTe$_2$ crystals were grown by the two-stage iodine transport~\cite{growth1}, that previously was successfully applied~\cite{growth1,growth2} for growth of other metal chalcogenides like NbS$_2$ and CrNb$_3$S$_6$. The WTe$_2$ composition is verified by energy-dispersive X-ray spectroscopy. The X-ray diffraction confirms $Pmn2_1$ orthorhombic single crystal WTe$_2$ with lattice parameters $a=3.48750(10)$~\AA, $b= 6.2672(2)$~\AA, and $c=14.0629(6)$~\AA. We check by standard magnetoresistance measurements that our WTe$_2$ crystals demonstrate large, non-saturating positive magnetoresistance in normal magnetic field, which goes to zero in parallel one,  as it has been shown for WTe$_2$ Weyl semimetal~\cite{MR}, see Ref.~\onlinecite{II-WSM3} for details of magnetoresistance measurements.

The single-crystal flakes of WTe$_2$  are obtained by regular mechanical exfoliation, also known as scotch-tape technique. Next, the exfoliated samples were transferred on the insulating SiO$_2$ substrate. While we need thick three-dimensional flakes to preserve  WTe$_2$ semimetal properties, we use two different techniques for Ohmic contacts fabrication for the flakes of different thickness. 

For the  thinnest, 300--600~nm flakes, the Au leads are defined over the flake surface by standard photolithography and lift-off technique after thermal evaporation of 70~nm Au, see the AFM image in  Fig.~\ref{sample}. As usual, thin flakes are about 10-30~$\mu$m in the lateral size, so only two or three Au leads can be placed over the flake to form Ohmic contacts with 5~$\mu$m distance (which should exceed $l_e\approx$1~$\mu$m, the mean free path in WTe$_2$). These samples are mostly suitable for the two-point transport measurements. 

The thicker (1-3~$\mu$m) flakes are about 100 $\mu$m in lateral size, which allows different multiple contact geometries. However, standard 70-100 nm thick Au leads can not be formed across the 1-3~$\mu$m  step, so we use different contact technique. Thick flakes are transferred to SiO$_2$ substrate with pre-defined Au  leads pattern, the flake is slightly pressed to the leads by another oxidized silicon substrate. Weak pressure is applied with a special metallic frame, which keeps the substrates strictly parallel. This procedure has been verified  to provide electrically stable contacts with high quality interfaces~\cite{II-WSM3,II-WSM4,cdas,timnal}. Also,  WTe$_2$ surface with Au contacts is protected from any contamination by SiO$_2$ substrate in this case.

Typical sample resistance is about 10--50~Ohm. For correct measurement of low-resistance samples one have to directly define current $I$  and measure the resulting voltage drop $V$. For the two-point measurements (small flakes), one of the contacts is grounded, dc current varies within $\pm$3~mA range at the neighbor contact, see  Fig.~\ref{sample}. To obtain $dV/dI(I)$ curves, the dc current $I$ is additionally modulated by a small ac component (0.03~mA)  at the 1600 Hz frequency. The ac voltage component is measured by lock-in, it is proportional to the differential resistance $dV/dI$ at a given $I$ value. We verify that the obtained $dV/dI(I)$ curves are independent of the modulation frequency in the range 1~kHz--10~kHz, which is determined by the applied filters. For large flakes we use a standard four-point scheme with two separate potential probes. 

This measurements can  be carried out in external electric field by applying gate voltage to the silicon wafer, separated from the flake by 300~nm SiO$_2$ layer. 
The measurements are  performed at room temperature for  WTe$_2$ samples of different thicknesses and lateral sizes, since ferroelectric domains have been previously  observed in WTe$_2$ semimetal at room temperature~\cite{WTe2_fer}.

\section{Experimental results}

\begin{figure}[t]
\center{\includegraphics[width=\columnwidth]{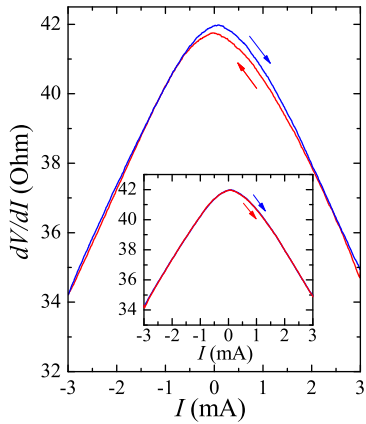}}
\caption{Hysteresis in the experimental $dV/dI(I)$ curves for the 400 nm thick WTe$_2$ sample for two current sweep directions.  Sweep directions for blue and red  curves  are shown by arrows of the same color. The $dV/dI(I)$ curves are affected by the sign of the current change, while inset demonstrates perfect reproducibility of the  curves, obtained for a single current sweep direction. Due to the measurement sequence and the lowest sweep rate (60 min for the full $\pm$3~mA sweep range), the hysteresis maximum is shifted to positive currents.}
\label{IV}
\end{figure}

Fig.~\ref{IV} shows the examples of experimental $dV/dI(I)$ curves for the thinnest, 400 nm sample from Fig.~\ref{sample}.   Differential resistance  $dV/dI(I)$ is a maximum (42~Ohm) at zero bias, it falls symmetrically at  positive and negative currents on about 20\% in a full current range. To our surprise,  we observe small but noticeable hysteresis in the experimental $dV/dI$ with current sweep direction, as demonstrated by red and blue curves. It is important, that $dV/dI(I)$ curves coincide perfectly if they are obtained for the same sweep direction, see the  inset to Fig.~\ref{IV}.  Thus, WTe$_2$ differential resistance is affected by the sign of the current change in the main field of Fig.~\ref{IV}.

The hysteresis is clearly  not symmetric  in Fig.~\ref{IV}, the maximum difference between the $dV/dI(I)$ curves  is shifted to positive currents due to extremely low sweep rate (60 min for the full $\pm$3~mA range). This well corresponds to the measurement procedure: the blue curve (current sweep from -3~mA to +3~mA) is obtained immediately after the red one (from +3~mA to -3~mA). 

\begin{figure}[t]
\center{\includegraphics[width=\columnwidth]{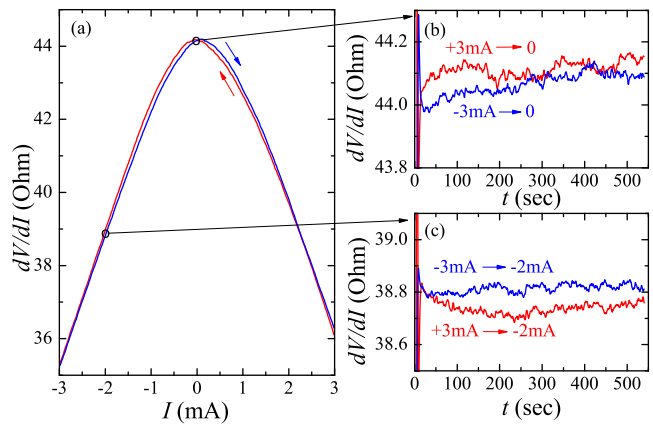}}
\caption{(a) $dV/dI(I)$ curves for the same 400 nm sample, as obtained at three times  higher sweep rate in comparison with Fig.~\ref{IV}. The hysteresis is symmetric and even more pronounced in this case. (b,c) Time-dependent relaxation curves $dV/dI(t)$ for two fixed current values $I=0$ and -2~mA in (b) and (c), respectively. For every panel (b) or (c), the curves are obtained after keeping the sample for 10 minutes at two different dwelling currents $\pm$3~mA. The relaxation curves $dV/dI(t)$ are clearly different in these cases, so $dV/dI$ depends on the sign of the current change.
}
\label{IV+relax}
\end{figure}

$dV/dI(I)$ curves are shown in Fig.~\ref{IV+relax} (a) for higher sweep rate (20 min for the full $\pm$3~mA range) for the same 400 nm sample. The hysteresis is even more pronounced in this case, so it reflects some  slow relaxation process in charge transport through  WTe$_2$.

To demonstrate the relaxation directly, we show $dV/dI(t)$ time-dependent curves for two different  current values $I=0$ and -2~mA, see  Fig.~\ref{IV+relax} (b) and (c). For a single pair of the curves, the sample is kept   at a fixed dwelling current  (3~mA or -3~mA) for 10~min. Current is abruptly set to the required  value (e.g. $I=0$ in (b)), time dependence of  $dV/dI$ is traced immediately after that. As a result, we observe slow relaxation in $dV/dI$, while there is clear difference between the curves for two dwelling currents $\pm$3~mA, see Fig.~\ref{IV+relax} (b)  and (c). Thus, there is time-dependent relaxation of $dV/dI$ differential resistance, which  depends on the sign of the current change. This behavior well correlates with the hysteresis in $dV/dI(I)$ curves in Figs.~\ref{IV} and Fig.~\ref{IV+relax} (a).

\begin{figure}[t]
\center{\includegraphics[width=\columnwidth]{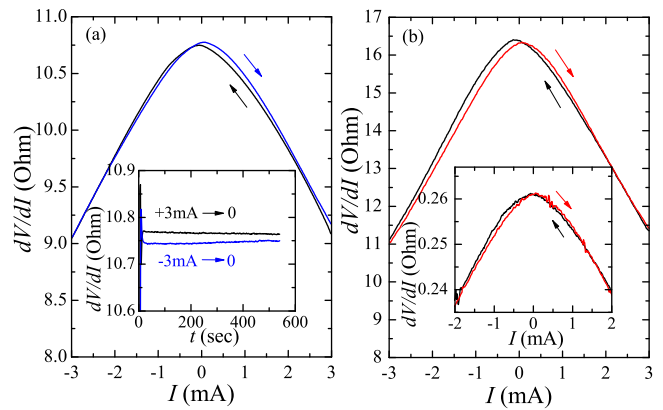}}
\caption{ Qualitatively similar $dV/dI(I)$ behavior for the 600~nm thick WTe$_2$ flake (a) and for the 3~$\mu$m one (b), the samples also differ by the contact preparation technique (see the main text). Relaxation is directly shown by time-dependent $dV/dI(t)$ curves  in the inset to (a). The overall $dV/dI(I)$ shape with hysteresis is also demonstrated in a standard four-point connection scheme in the inset to (b), which excludes Au-WTe$_2$ interfaces as a possible origin of the effect.}
\label{IV two scheme}
\end{figure}

These results can be reproduced for samples of different thicknesses and lateral sizes, and, therefore, of different contact preparation  techniques, see Fig.~\ref{IV two scheme}.  $dV/dI(I)$ curves are shown in Fig.~\ref{IV two scheme} (a) for the 600~nm thick  sample, the Au leads are still evaporated over the WTe$_2$ flake. The overall sample behavior is quite similar to one in Figs.~\ref{IV} and~\ref{IV+relax} (a): $dV/dI(I)$ curves also depend on the current sweep direction, there is a well-defined difference between the $dV/dI(t)$ relaxation curves at $I=0$ for two dwelling currents $\pm$3~mA, as depicted in the inset. 
Similar results can be obtained for the 3~$\mu$m thick WTe$_2$ flake, which is pressed to the pre-defined Au leads pattern, see Fig.~\ref{IV two scheme} (b). 

These effects  are defined by bulk WTe$_2$ properties, which can be confirmed by measurements in a standard four-point connection scheme, see the inset to Fig.~\ref{IV two scheme} (b) for the large-area 3~$\mu$m thick WTe$_2$ flake. In this case,  Au-WTe$_2$ interfaces are excluded, which leads to  much lower $dV/dI(I)$ values. Nevertheless, both the overall $dV/dI(I)$ shape and the dependence on the current sweep direction are well reproduced in the inset.

\begin{figure}[t]
\center{\includegraphics[width=\columnwidth]{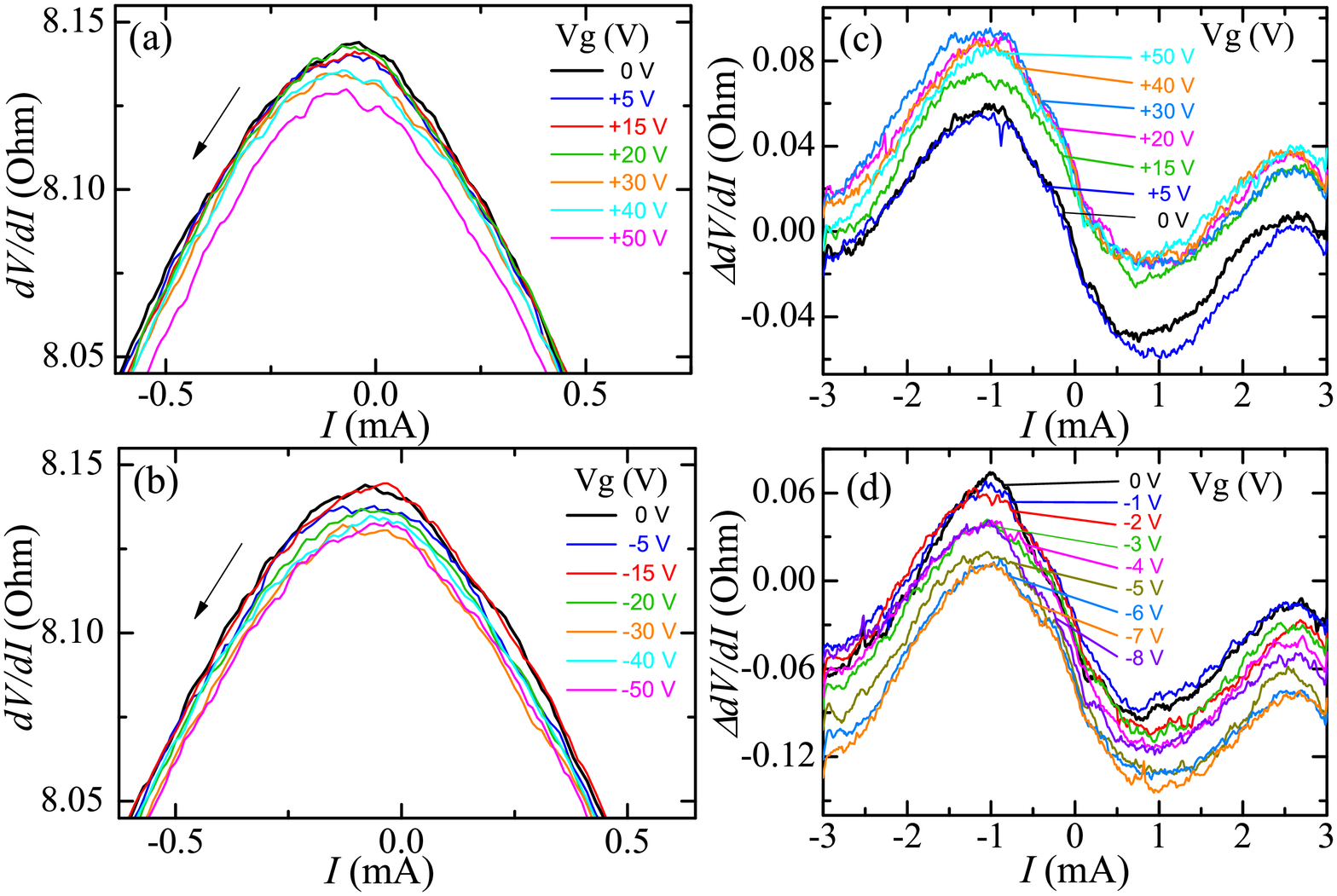}}
\caption{ (a) and (b) Gate-voltage dependence of  $dV/dI(I)$, obtained for a single current sweep direction for the 4~$\mu$m thick sample. The curves are shifted symmetrically for two different gate voltage polarities, which contradicts to the standard asymmetric field-effect transistor behavior. There is no gate leakage in the present gate voltage range.  (c) and (d): Hysteresis amplitude  $\Delta dV/dI(I)$, obtained by subtracting two curves for opposite current sweep direction at fixed gate voltage. The $\Delta dV/dI(I)$ curves are shown for the 4~$\mu$m thick sample in (c) and for the thinner, 600~nm one (d).  The curves are even quantitatively similar for different samples, they are of odd behavior with a maximum at the -1~mA negative current and a minimum at the +1~mA positive one. There is no any systematic dependence on $V_g$ for $\Delta dV/dI(I)$ in (d).}
\label{IV and Delta IV at Vg}
\end{figure}

We can also study effect of the normal-to-the-plane electric field on $dV/dI(I)$ curves by using silicon substrate as a gate electrode. We check that for the 300~nm SiO$_2$ thickness, there is no gate leakage at least in the $\pm$50~V range. Experimental $dV/dI(I)$ curves are shown for a single current sweep direction in  Fig.~\ref{IV and Delta IV at Vg} (a) and (b) for two gate voltage polarities, respectively. Increasing the gate voltage value shifts $dV/dI(I)$ curves down irrespective of the gate voltage sign, in contrast to the standard asymmetric field-effect transistor behavior.  

The  hysteresis amplitude can be demonstrated directly by subtracting two $dV/dI(I)$ curves for opposite current sweep directions at fixed gate voltage. The result is shown as $\Delta dV/dI(I)$ in Fig.~\ref{IV and Delta IV at Vg}  for 4~$\mu$m thick flake with pre-defined contacts (c), and for the 600~nm one with evaporated contacts (d). The curves are even quantitatively similar in (c) and (d), they are of odd behavior with a $\Delta dV/dI(I)$ maximum at the -1~mA negative current and a $\Delta dV/dI(I)$ minimum at the +1~mA positive one. The curves are shifted vertically for different gate voltages, but one can not see any systematic dependence on $V_g$, in contrast to a single-sweep $dV/dI(I)$ curves in Fig.~\ref{IV and Delta IV at Vg} (a) and (b).

\section{Discussion}

All five WTe$_2$ samples demonstrate   symmetric decrease in $dV/dI(I)$, which should be expected for zero-gap semiconductors. In this case, even low electrochemical potential shift leads to the noticeable change of the carrier concentration even away from the charge neutrality point. Electrochemical potential difference appears between the potential contacts at finite $I$, so the decrease in $dV/dI(I)$ reflects the increased concentration near the high-potential probe. The effect is symmetric due to the overall symmetry of the sample. 

On the other hand, one can not expect the observed dependence of $dV/dI$ differential resistance on the sign of the current change. Also, the symmetric $dV/dI(I)$ gate voltage shift for both $V_g$ signs is not consistent with a standard accumulation/depletion field effect in semiconductors in Fig.~\ref{IV and Delta IV at Vg} (a) and (b).

For all five WTe$_2$ samples, $dV/dI(I)$s demonstrate smooth behavior, there are no special points which could be associated with phase transitions due to the Joule heating of the sample. It well corresponds to the known WTe$_2$ properties, since WTe$_2$ crystal structure (Td) is known to be  stable in a wide temperature range~\cite{WTe2_str1,WTe2_str2,WTe2_str3} at ambient pressure. Phase transitions can only be observed in WTe$_2$, subjected to high pressures~\cite{WTe2_PhT1,WTe2_PhT2,WTe2_PhT3}, so one can not associate the observed hysteresis in $dV/dI(I)$ with any phase-change effects. 

Also, Au-WTe$_2$ interfaces have no contribution to the hysteresis in $dV/dI(I)$, since it can be reproduced by the four-point measurements in the inset to Fig.~\ref{IV two scheme} (b). Thus, the observed effects can not be connected with previously reported~\cite{NVRS1}  gold adatom absorption at defects. 

Surprisingly, even well-conducting WTe$_2$ single crystals  demonstrate  ferroelectricity at room temperature, which has been shown by direct visualization of ferroelectric domains~\cite{WTe2_fer}. The spontaneous polarization of these domains is normal to the WTe$_2$ layers, it can be affected by external electric field~\cite{WTe2_fer}.  This is a natural mechanism for the observed effects in $dV/dI$ both for the driving current and for the gate voltage behavior: 

(i) The hysteresis amplitude  $\Delta dV/dI(I)$ in Fig.~\ref{IV and Delta IV at Vg} (c) and (d) is identical to the first derivative of a standard ferroelectric hysteresis loop. Indeed, the dc circuit of our experimental setup is equivalent to so called Sawyer-Tower's circuit, which is usual for  polarization hysteresis measurements for ferroelectric films~\cite{ferr_book,scheme}. Due to the modulation technique, we obtain the first derivative of a standard  $I-V$ ferroelectric loop, which leads to the maximum at the negative current  (-1~mA) and the minimum at the positive one (+1~mA). The ferroelectric hysteresis originates from the polarization current: any change in the applied bias leads to the corresponding change in the crystal lattice deformation; the latter is responsible for the slow relaxation in $dV/dI(t)$ characteristics and the $dV/dI(I)$ dependence  on the sign of the current change.  The ferroelectric hysteresis originates from the ferroelectric spontaneous polarization: any change in the applied bias leads to the crystal lattice deformation and to the corresponding change in the polarization charge; the latter is responsible for the slow relaxation in $dV/dI(t)$ characteristics and the $dV/dI(I)$ dependence on the sign of the current change.

(ii) The achievable values of the gate electric field ($\sim 10^{6}$~V/m) are too small to align polarization of the whole WTe$_2$ flake, so the gate field can only affect the positions of the domain walls. At fixed $V_g$, it gives no contribution to the polarization current, so  $\Delta dV/dI(I)$ ferroelectric hysteresis loop is nearly independent of the gate voltage in  Fig.~\ref{IV and Delta IV at Vg} (c) and (d). Instead,  there should be a constant $dV/dI$ shift at a fixed gate voltage value, due to the strong coupling betweeen the carriers concentration and the electrochemical potential position in zero-gap  WTe$_2$. Thus, the observed symmetry in the gate voltage behavior of $dV/dI(V)$ in Fig.~\ref{IV and Delta IV at Vg} (a) and (b) reflects the  overall symmetry in $dV/dI(I)$ curves. 

The observed effects are nearly independent  of the sample thickness in Figs.~\ref{IV},~\ref{IV+relax}, and~\ref{IV two scheme}, which should be connected with the planar experimental geometry. The current is mostly concentrated  near the surface between the Au leads, where the ferroelectric domains have been observed previously by scanning techniques~\cite{WTe2_fer}.

\section{Conclusion}

In conclusion, we investigate electron transport along the surface of WTe$_2$ three-dimensional single crystals, which are characterized by coexistence of Weyl semimetal conductivity and ferroelectricity at room temperature. We find that non-linear behavior of $dV/dI(I)$ WTe$_2$ differential resistance is accompanied by slow relaxation process, which appears as the $dV/dI(I)$ dependence on the sign of the current change. This observation is confirmed by direct investigation of time-dependent relaxation curves. While strongly non-linear  differential resistance  should be expected for the zero-gap WTe$_2$, the slow relaxation  in  transport is very unusual for well-conducting semimetals  at room temperature. We establish, that non-monotonous dependence of the amplitude of the effect on driving current $\Delta dV/dI(I)$  well corresponds to  the known Sawyer-Tower's  ferroelectric hysteresis loop. This conclusion is also confirmed by gate voltage dependencies, so our results can be understood as a direct demonstration of WTe$_2$ ferroelectric polarization in charge transport experiment.

\section{Acknowledgement}

The authors are grateful to V.T. Dolgopolov for fruitful discussions and S. S. Khasanov for XPS, and x-ray tungsten ditelluride characterization. We gratefully acknowledge financial support  by  RF State task.

\end{document}